\documentclass[conference]{IEEEtran}
\IEEEoverridecommandlockouts
\usepackage[pdftex]{graphicx}
\usepackage[ruled]{algorithm2e}
\usepackage{algorithmic}
\usepackage[small]{caption}
\usepackage{float}
\usepackage{xspace}
\usepackage{fancyhdr}
\usepackage{amssymb,amsmath}
\usepackage{subfig}
\usepackage{epstopdf}
\usepackage{bbding}
\usepackage{cite} 
\usepackage{color}
\begin{document}

\title{Backscatter-Aided
NOMA V2X Communication under Channel Estimation Errors}
\author{Wali Ullah Khan$^1$, Muhammad Ali Jamshed$^2$, Asad Mahmood$^1$, Eva Lagunas$^1$, \\ Symeon Chatzinotas$^1$, Bj\"orn Ottersten$^1$ \\$^1$Interdisciplinary Centre for Security, Reliability and Trust (SnT), University of Luxembourg\\
$^2$James Watt School of Engineering, University of Glasgow, UK\\
\{waliullah.khan, asad.mahmood, eva.lagunas, symeon.chatzinotas, bjorn.ottersten\}@uni.lu\\
muhammadali.jamshed@glasgow.ac.uk\thanks{This work was supported by Luxembourg National Research Fund (FNR) under the CORE project RISOTTI C20/IS/14773976.}

}%

\maketitle

\begin{abstract}
Backscatter communications (BC) has emerged as a promising technology for providing low-powered transmissions in nextG (i.e., beyond 5G) wireless networks. The fundamental idea of BC is the possibility of communications among wireless devices by using the existing ambient radio frequency signals. Non-orthogonal multiple access (NOMA) has recently attracted significant attention due to its high spectral efficiency and massive connectivity. This paper proposes a new optimization framework to minimize total transmit power of BC-NOMA cooperative vehicle-to-everything networks (V2XneT) while ensuring the quality of services. More specifically, the base station (BS) transmits a superimposed signal to its associated roadside units (RSUs) in the first time slot. Then the RSUs transmit the superimposed signal to their serving vehicles in the second time slot exploiting decode and forward protocol. A backscatter device (BD) in the coverage area of RSU also receives the superimposed signal and reflect it towards vehicles by modulating own information. Thus, the objective is to simultaneously optimize the transmit power of BS and RSUs along with reflection coefficient of BDs under perfect and imperfect channel state information. The problem of energy efficiency is formulated as non-convex and coupled on multiple optimization variables which makes it very complex and hard to solve. Therefore, we first transform and decouple the original problem into two sub-problems and then employ iterative sub-gradient method to obtain an efficient solution. Simulation results demonstrate that the proposed BC-NOMA V2XneT provides high energy efficiency than the conventional NOMA V2XneT without BC.
\end{abstract}


\section{Introduction}
The nextG (i.e., beyond 5G) transportation systems are expected to improve traffic control, traffic efficiency, reliability, and passenger safety \cite{9521550}. Vehicle-to-everything (V2X) communication is one of the emerging technologies to make these applications available \cite{liu20206g}. The V2X networks (V2XneT) include vehicle-to-vehicle (V2V), vehicle-to-infrastructure (V2I), vehicle-to-roadside unit (V2R), vehicle-to-pedestrian (V2P), vehicle-to-unmanned-aerial-vehicle (V2U), and vehicle-to-satellite (V2S) \cite{9217500}. Thus, a connected network in the form of internet-of-vehicles (IoV) is formed which must support high reliability, massive connectivity, while providing energy efficient communications and low transmission latency \cite{wang2021green}. It is also important to mention here that the 3rd generation partnership project (3GPP) is already working on the V2XneT solutions for public safety \cite{9530506}. However, one of the critical issues in existing V2XneT is the use of the conventional orthogonal multiple access (OMA) technique that can only accommodate a limited number of vehicle connections on the available frequency spectrum. This scarcity of available frequency spectrum can cause traffic and data congestion in dense V2XneT. To address this issue, researchers in academia and industry are exploring new air interface techniques.

Recently, non-orthogonal multiple access (NOMA) has been
recognized as one of the promising air interface techniques for 6G wireless networks due to high spectral efficiency and massive connectivity \cite{8861078}. The important feature of NOMA is the use of the power-domain for multiple access, compared to previous air interface techniques that relied on other domains, i.e., time, frequency, and code \cite{liu2021application}. In NOMA, multiple vehicles can access the same spectrum/time resource for communications. More specifically, multiple vehicles can be multiplexed through different power levels by using the superposition coding technique at the transmitter side. Then the vehicle with strong channel conditions can apply SIC and decode its desired signal \cite{9411862}. In NOMA communications, by allocating more power, the quality of services of vehicles with weak channel conditions can be guaranteed.

With the increase in the number of vehicles and the requirements of high data
rates, energy consumption increases significantly. This is very problematic as one of the main issues in green V2XneT is the increasing levels of CO2 emissions \cite{8214104}. Thus, there has been increased interest to use ambient radio frequency (RF) waves for communications between different vehicles. In this regard, a promising technology is backscatter communications (BC) which can ease the situation \cite{9261963}. Using existing RF resources, BC can transfer the data between different vehicles in the network. Therefore, BC can extend the lifetime of devices (that act as tags) by reflecting the existing RF signals towards intended vehicles without exploiting any oscillatory circuity. The performance of BC in different scenarios using conventional OMA technique has been extensively studied in literature \cite{9348943,9363336,khan2021learning,9162720,9507551,9024401,tran2020throughput}. 
\begin{figure}[t]
\centering
\includegraphics [width=0.45\textwidth]{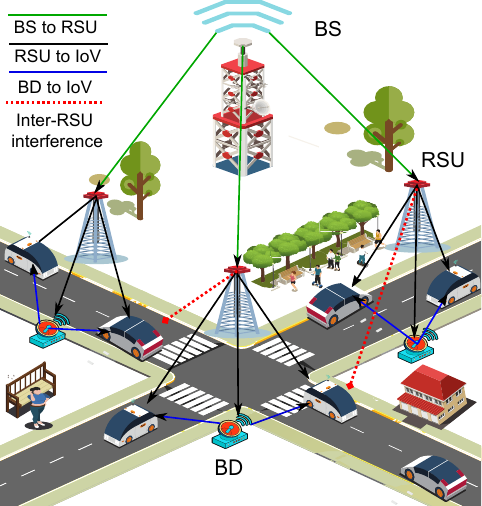}
\caption{System Model of BC-NOMA V2X communications.}
\label{blocky}
\end{figure}

Researchers in industry and academia have recently conducted studies on resource optimization of BC-NOMA networks. For example, Khan {\em et al.} \cite{9543581} have provided an energy-efficient optimization framework for multi-cell BC-NOMA IoV networks under the assumption of imperfect SIC decoding error. To improve the minimum throughput, the work in \cite{liao2020resource} has proposed a resource allocation framework for optimizing power, time and reflection coefficient in full-duplex BC-NOMA networks. Besides that, the work of \cite{9223730} has presented the optimization of power and reflection coefficient for green communications in BC-NOMA systems. To enhance spectral efficiency, the authors of \cite{khan2021integration} have explored a new problem of resource management in multi-cell BC-NOMA networks under the assumption of SIC decoding error. Moreover, to minimize interference between uplink and downlink transmission, Ding {\em et al.} \cite{9420716} have investigated the achievable sum rate maximization problem for the BC-NOMA system. The problem of physical layer security is also investigated in \cite{khan2021joint} for the BC-NOMA system, where the transmit power and reflection coefficient have jointly optimized to maximize the sum secrecy rate. Of late, the problem of improving energy efficiency is investigated by Ahmed {\em et al.} \cite{ahmed2021backscatter} for multi-cell BC-NOMA network to simultaneously optimize the transmit power and reflection coefficient under SIC decoding error.

Although many works have been carried out on BC-NOMA networks, the existing works have considered perfect channel state information (CSI) in their system models, which are not practical in real scenarios. To the best of our knowledge, the problem of energy-efficient communication for BC-NOMA cooperative V2XneT under imperfect CSI has not been investigated previously. Therefore, this paper investigates a new problem to simultaneously optimize the transmit power of base station (BS), roadside units (RSUs), and reflection coefficient of backscatter devices (BDs) under perfect and imperfect CSI. To solve the optimization problem efficiently, we first transform and decouple the original problem into two sub-problem and then adopt the iterative sub-gradient method. Simulation results are also provided to demonstrate the effect of channel estimation error and the benefits of the proposed optimization framework against the benchmark optimization. The rest of the paper is structured as: the second section studies the system model and problem formulation; the third section presents the proposed optimization solution; the fourth section discusses the simulations results; the fifth section concludes this paper.

\section{System Model and Problem Formulation}
Consider a BC-NOMA cooperative V2XneT, as illustrated in Figure \ref{blocky}. In the proposed model, we consider a single BS and multiple RSUs, where each RSU accommodates two vehicles using decode-and-forward protocol. We also consider a BD in the coverage area of RSUs. The direct link from BS to vehicles is missing due to high shadowing. In this work, it is assumed that: i) All the devices in the network are equipped with single antenna; ii) The CSI over all links is imperfect due to the channel estimation errors; iii) The wireless channels are independent and undergo Rayleigh fading. This model considers half duplex communication, where the transmission process is completed in two different time slots. In the first time slot, a BS sends the superimposed signal to its serving RSUs utilizing NOMA protocol. In the second time slot, RSUs first decode the superimposed signal and then forward it to its serving vehicles.

\subsection{First Time Slot (transmission from BS to RSUs)}
In the first time slot, transmission between BS and RSUs takes place. Although we consider two RSUs for simplicity, our model can be easily extended to multiple RSUs, which will be investigated in our future study. A superimposed signal that BS transmits (denoted as $x$) can be expressed as  
\begin{align}
x=\sum\limits_{m=1}^2\sqrt{P\alpha_m}x_m,
\end{align}
where $P$ is the transmit power of BS and $\alpha_m$ denotes the power allocation coefficient of the RSU (denoted as $R_m$), for $m\in\{1,2\}$. Moreover, $x_m$ represents the unit power signal of $R_m$. The channel from BS to $R_m$ can be modeled as $h_m=H_m\times d_m^{\frac{-\zeta}{2}}$, where $H_m$ is the Rayleigh fading coefficient, $d_m$ denotes the distance from BS to $R_m$ and $\zeta$ represents the path-loss exponent \cite{8540884}. In this work, we consider the error in channel estimation, hence the CSI is imperfect. Using the minimum mean square error (MMSE) model, the channel of $R_m$ from the BS is estimated as $h_m=\hat h_m+\epsilon_m$, where $\hat h_m$ is the estimated channel gains of $h_m$ with variance $\sigma^2_{\hat h_m}$ and $\epsilon_m$ represents the estimated channel error with zero mean and $\sigma^2_{\epsilon_m}$ variance. For the convenience of discussion, the case of constant estimation error ($\sigma^2_{\epsilon_m}=\sigma^2_{\epsilon}$) for all channels is considered in this work. It is important to note that both $\hat h_m$ and $\epsilon_m$ are uncorrelated. The signal that $R_m$ receives from BS can be expressed as  
\begin{align}
y_m=\hat h_mx+\epsilon x+\varpi_m,\label{ym},
\end{align}
where $\varpi_m$ is the additive white Gaussian noise (AWGN) with zero mean and $\sigma^2$ variance. We assume that the channel gains of RSUs are arranged as $\hat h_{1}>\hat h_2$. Therefore, $R_1$ can apply SIC to decode its signal while $R_2$ cannot apply SIC and decode its signal by treating the signal of $R_1$ as a noise. Based on these observations, the data rate $C_1$ and $C_2$ can be written as $C_{1}=t_1B\log_2(1+\gamma_1)$ and $C_{2}=t_1B\log_2(1+\gamma_2)$, where $t_1$ shows the first transmission slot which should be equal to 1/2. The terms $\gamma_1$ and $\gamma_2$ are the signal-to-interference-plus-noise-ratio (SINR) which can be stated as
\begin{align}
\gamma_1 & =\frac{|\hat h_1|^2P\alpha_1}{P\sigma^2_\epsilon(\alpha_1+\alpha_2)+\sigma^2}.\\
\gamma_2 & =\frac{|\hat h_2|^2P\alpha_2}{|\hat h_2|^2P\alpha_1+P\sigma^2_\epsilon(\alpha_1+\alpha_2)+\sigma^2}.
\end{align}

\subsection{Second Time Slot (transmission from RSUs to vehicles)}
In this time slot, transmission between RSUs and vehicles takes place. The RSUs first regenerate the superimposed signal and then forward it. The signal that $R_m$ transmits (denoted as $s_m$) can be written as 
\begin{align}
s_m=\sum\limits_{i=1}^2\sqrt{Q_m\beta_{i,m}}s_{i,m},
\end{align}
where $Q_m$ is the transmit power of $R_m$, $\beta_{i,m}$ represents the power allocation coefficients of the vehicle (denoted as $V_{i,m}$), for $i\in\{1,2\}$ and $s_{i,m}$ denotes the unit power signals of $V_{i,m}$. The channels used in this time slot can be modeled and estimated similar to the first time slot. However, for the simplicity, we denote the channel from $R_m$ to $V_{i,m}$ as $\hat h_{i,m}$. Without loss of generality, we assume that the channel gains of $V_{1,m}$ is stronger than $V_{2,m}$, i.e., $|\hat h_{1,m}|^2>|\hat h_{2,m}|^2$.

During the transmission in the second time slot, a BD in the geographical area of $R_m$ (stated as $B_{m}$) also receives the superimposed signal $s_m$ from $R_m$. $B_{m}$ first harvests energy from $s_m$, then modulate its own message $z_{m}$ and reflect it towards $V_{i,m}$, where $\mathbb E[|z_m|^2]=1$ and $\mathbb E[.]$ represents the expectation operation. Since we consider imperfect CSI, therefore the signal that $V_{i,m}$ and receives from $R_m$ can be expressed as
\begin{align}
y_{i,m}&=\hat h_{i,m}s_m+\sqrt{\xi_m}\hat h^{b}_{i,m}\hat h_{b,m}s_mz_m\nonumber\\&+\epsilon s_m+\epsilon\sqrt{\xi_m}+\varpi_{i,m}, \label{y_{i,m}}
\end{align}
where $\xi_m$ is the reflection coefficient of $B_m$ and $\hat h^{b}_{i,m}$ denotes the channel gain between $B_m$ and $V_{i,m}$. Further, $\hat h_{b,m}$ represents the channel gain between $R_m$ and $B_m$ while $\varpi_{i,m}$ states the AWGN. Based on the received signal in (\ref{y_{i,m}}), the data rate of $V_{1,m}$ and $V_{2,m}$ cab be formulated as $C_{1,m}=t_2B\log_2(1+\gamma_{1,m})$ and $C_{2,m}=t_2B\log_2(1+\gamma_{2,m})$, where $t_2$ shows second slot, $\gamma_{1,m}$ and $\gamma_{2,m}$ are the SINRs as
\begin{align}
\gamma_{1,m} & =\frac{Q_m\beta_{1,m}(|\hat h_{1,m}|^2+\xi_m|\hat h^{b}_{1,m}|^2|\hat h_{b,m}|^2)}{\sigma^2_\epsilon(Q_m(\beta_{1,m}+\beta_{2,m})+\xi_m)+I^{m'}_{1,m}+\sigma^2}.\\
\gamma_{2,m} & =\frac{Q_m\beta_{2,m}(|\hat h_{2,m}|^2+\xi_m|\hat h^{b}_{2,m}|^2|\hat h_{b,m}|^2)}{\varPi_{2,m}+\sigma^2_\epsilon(Q_m(\beta_{1,m}+\beta_{2,m})+\xi_m)+I^{m'}_{2,m}+\sigma^2}.
\end{align}
where $\varOmega_{1,m}=\xi_m|\hat h_{1,m}^b|^2|\hat h_{b,m}|^2$ and $\varOmega_{2,m}=\xi_m|\hat h_{2,m}^b|^2|\hat h_{b,m}|^2$ refer to the useful received from the BDs. Furthermore, $\varPi_{2,m}=Q_m\beta_{1,m}(|\hat h_{2,m}|^2+\xi\hat h^{b}_{2,m}\hat h_{b,m})$ is the interference due to NOMA transmission, and $I^{m'}_{i,m}=|\hat h^{m'}_{i,m}|^2Q_{m'}$ is the co-channel interference among RSUs. According to the decode and forward protocol at RSUs, the end-to-end rate can be calculated as \cite{8678396}
\begin{align}
\bar C=\frac{1}{2}\text{min}\{C_m,C_{i,m}\}.
\end{align}
Then, the sum rate of V2XneT can be expressed as
\begin{align}
\bar C_{sum}=\sum\limits_{m=1}^2\sum\limits_{i=1}^2\frac{1}{2}\text{min}\{C_m,C_{i,m}\}.
\end{align}

\subsection{Problem Formulation}
The objective of this work is to provide energy-efficient communication in BC-NOMA V2XneT by optimization the transmit power of BS, RSUs, and reflection coefficient of BDs. It can be achieved by formulating the following optimization problem 
\begin{alignat}{2}
\text{(P)}&\ \underset{{\alpha_{m},\beta_{i,m},\xi_m}}{\text{min}} \bigg\{\sum\limits_{m=1}^2P\alpha_m+\sum\limits_{m=1}^2\sum\limits_{i=1}^2Q_m\beta_{i,m}\bigg\}\nonumber\\
s.t. &\  (A1): \sum\limits_{m=1}^2C_m\geq C_{min}, \nonumber\\
&\ (A2): \sum\limits_{i=1}^2C_{i,m}\geq C_{min}, m\in\{1,2\}, \nonumber\\
&\ (A3): \sum\limits_{m=1}^2P\alpha_m\leq P_{max},\nonumber\\
&\ (A4): \sum\limits_{m=1}^2\alpha_m\leq 1, \nonumber\\
&\ (A5): \sum\limits_{i=1}^2Q_m\beta_{i,m}\leq Q_{max}, m\in\{1,2\}, \nonumber\\
&\ (A6): \sum\limits_{i=1}^2\beta_{i,m}\leq 1, m\in\{1,2\},\nonumber\\
&\ (A7): 0\leq\xi_m\leq1, m\in\{1,2\},\label{19}
\end{alignat}
where the objective of (P) is to minimize the total transmit power of the BC-NOMA V2XneT. Constraints $(A1)$ and $(A2)$ ensure the minimum rate and first and second slots, where $C_{min}$ shows the minimum rate threshold. Constraints $(A3)$ and $(A5)$ limit the transmit power of BS and RSUs, where $P_{max}$ and $Q_{max}$ are the maximum power budget that BS and RSU. Constraints $(A4)$ and $(A6)$ describe the power allocation according to NOMA protocol. Constraint $(A7)$ controls the reflection coefficient of BD. 

\section{Proposed Solution}
The problem (P) is coupled on multiple variables which makes it very complex and hard to solve. Thus, we first transform and decouple it into two sub-problems, i.e., i) power allocation at BS; 2) power allocation at RSUs and reflection coefficient at BDs. Then, we adopt iterative sub-gradient method to obtain sub-optimal yet efficient solution \cite{boyd2004convex}. For any given value of transmit power at RSUs and reflection coefficient at BDs, the power minimization problem in (\ref{19}) can be simplified as:
\begin{alignat}{2}
&\ \underset{{(\alpha_{1},\alpha_{2})}}{\text{min}}\ P(\alpha_1+\alpha_{2})\label{14}\\
s.t. &\  (A8):  |\hat h_1|^2P\alpha_1\geq(2^{C_{min}}-1)(P\sigma^2_\epsilon(\alpha_1+\alpha_2)+\sigma^2), \nonumber\\
& \ (A9): |\hat h_2|^2P\alpha_2\geq(2^{C_{min}}-1)\times\nonumber\\&\quad\quad\quad\ (|\hat h_2|^2P\alpha_1+P\sigma^2_\epsilon(\alpha_1+\alpha_2)+\sigma^2), \nonumber\\
& \ (A10): P(\alpha_1+\alpha_2)\leq P_{max},\nonumber\\
& \ (A11): \alpha_1+\alpha_2\leq 1,\nonumber
\end{alignat}
where the objective in (\ref{14}) is to minimize the total BS transmit power. Constraint $(A8)$ and constraint $(A9)$ guarantee the minimum rate of $R_1$ and $R_2$, respectively. Constraint $(A10)$ limits the transmit power of BS while constraint $(A11)$ is the power allocation limit of $R_1$ and $R_2$. To solve (\ref{14}), we employ a sub-gradient method, in which we first define the Lagrangian function and then compute its derivation with respect to $\alpha_1$ and $\alpha_2$. Here we omit the detail derivation for sake of simplicity and limited space, the value of $\alpha^*_1$ and $\alpha^*_2$ can be expressed as   
\begin{align}
\alpha^*_1 & = P+\lambda_1 P+(2^{C_{min}}-1)\psi_2 (|\hat{h}_2|^2 P+P \sigma_{\epsilon}^2)\nonumber\\&+\psi_1 ((2^{C_{min}}-1) P \sigma_{\epsilon}^2-|\hat{h}_1|^2 P)+\lambda_2\alpha_2\\
\alpha^*_2 & = P+\lambda_1 P+(2^{C_{min}}-1) P \psi_1 \sigma_{\epsilon}^2\nonumber\\&+\psi_2 ((2^{C_{min}}-1) P \sigma_{\epsilon}^2-|\hat{h}_2|^2 P)+\lambda_2\alpha_1
\end{align}
where $\psi_1,\psi_2,\lambda_1, \lambda_2$ are the Lagrangian multipliers.
Next we iteratively update $\alpha_1$, $\alpha_2$ and the Lagrangian multipliers as
\begin{align}
\alpha_1^{(itr+1)}&=\big(\alpha_1(itr)- \delta(itr) \alpha^*_1\big)^+\\
\alpha_2^{(itr+1)}&=\big(\alpha_2(itr)- \delta(itr) \alpha_2^*\big)^+\\
\psi_1^{(itr+1)} & =\big(\psi_1(itr)-\delta(itr)(|\hat h_1|^2P\alpha_1\nonumber\\&-(2^{C_{min}}-1)P\sigma_\epsilon^2(\alpha_1+\alpha_2)+\sigma^2)\big)^+\\
\psi_2^{(itr+1)} & =\big(\psi_2(itr)-\delta(itr)(|\hat h_2|^2P\alpha_2-(2^{C_{min}}-1)\nonumber\\&(|\hat h_2|^2P\alpha_1+P\sigma_\epsilon^2(\alpha_1+\alpha_2))+\sigma^2)\big)^+\\
\lambda_1^{(itr+1)}&=\big(\lambda_1(itr)-\delta(itr)(P_{max}-P(\alpha_1+\alpha_2))\big)^+\\
\lambda_2^{(itr+1)}&=\big(\lambda_2(itr)-\delta(itr)(1-(\alpha_1+\alpha_2))\big)^+
\end{align}
where $itr$ shows the iteration index and $\delta$ is the nonnegative step size. The above iterative process will continue until the required criterion is satisfied. 

Accordingly, for a given transmit power at BS, the problem of power allocation at RSUs and reflection coefficient of BDs can be formulated as
\begin{alignat}{2}
 & \underset{{(\beta_{1,m},\beta_{2,m},\xi_{m})}}{\text{min}}\sum\limits_{m=1}^2Q_m(\beta_{1,m}+\beta_{2,m})\label{23}\\
s.t. &\  (A12):  Q_{m}\beta_{1,m}(|\hat h_{1,m}|^2+\varOmega_{1,m})\geq(2^{C_{min}}-1)\nonumber\\&\ (\sigma^2_\epsilon(Q_m(\beta_{1,m}+\beta_{2,m})+\xi_m)+I_{1,m}^{m'}+\sigma^2),m\in\{1,2\}, \nonumber\\
& \ (A13): Q_{m}\beta_{2,m}(|\hat h_{2,m}|^2+\varOmega_{2,m})\geq(2^{C_{min}}-1)(\varPi_{2,m}\nonumber\\&+\sigma^2_\epsilon(Q_m(\beta_{1,m}+\beta_{2,m})+\xi_m)+I_{2,m}^{m'}+\sigma^2),m\in\{1,2\}, \nonumber\\
& \ (A14): Q_m(\beta_{1,m}+\beta_{2,m})\leq Q_{max},m\in\{1,2\},\nonumber\\
& \ (A15): \beta_{1,m}+\beta_{2,m}\leq 1,m\in\{1,2\},\nonumber\\
& \ (A16): 0\leq\xi_m\leq 1,m\in\{1,2\},\nonumber
\end{alignat}
where constraints $(A12)$ and $(A13)$ ensure the minimum rate of $V_{1,m}$ and $V_{2,m}$, respectively. Constraint $(A14)$ and constraint $(A15)$ control the transmit power of RSUs according to the NOMA protocol while $(A16)$ is the reflection coefficient constraint. Similar to (\ref{14}), we exploit the sub-gradient method to obtain the efficient solution. After necessary calculation, the value of $\beta^*_1$ $\beta^*_2$, and $\xi_{m}$ can be stated as
\begin{align}
\beta^*_{1,m}&= Q_m(1-|\hat{h}_{1,m}|^2 \eta_{1,m}-\xi_{m} |\hat{h}_{1,m}^b|^2 |\hat{h}_{b,m}|^2 \eta_{1,m}\nonumber\\&+(2^C_{min}-1)(\eta_{1,m}+\eta_{2})\sigma_{\epsilon}^2+ \mu_{m})\\
\beta^*_{2,m}&= Q_m(1-|\hat{h}_{2,m}|^2 \eta_{2,m}-\xi_{m} |\hat{h}_{2,m}^b|^2 |\hat{h}_{b,m}|^2 \eta_{2,m}\nonumber\\&+(2^C_{min}-1)(\eta_{1,m}+\eta_{2,m})\sigma_{\epsilon}^2+ \mu_{m})\\
\xi^*_{m}&=-\beta_{1,m}|\hat{h}_{1,m}^b|^2 |\hat{h}_{b,m}|^2 \eta_{1,m} Q_m-\beta_{2,m}|\hat{h}_{2,m}^b|^2 |\hat{h}_{b,m}|^2 \nonumber\\&\times\eta_{2,m} Q_m+(2^C_{min}-1)(\eta_{1}+\eta_{2,m})\sigma_{\epsilon}+\upsilon_m
\end{align}
where $\eta_{1,m},\eta_{2,m},\mu_{m},\zeta_m,\upsilon_m$ are the Lagrangian multipliers. Now we iteratively update $\beta^*_1$ $\beta^*_2$, and $\xi_{m}$ along with the Lagrangian multipliers as
\begin{align}
\beta_{1,m}^{(itr+1)}&=\big(\beta_{1,m}(itr)- \delta(itr) \beta^*_{1,m}\big)^+\\
\beta_{2,m}^{(itr+1)}&=\big(\beta_{2,m}(itr)- \delta(itr) \beta^*_{2,m}\big)^+\\
\xi_{m}^{(itr+1)}&=\big(\xi_{m}(itr)- \delta(itr) \xi^*_{m}\big)^+\\
\eta_{1,m}^{(itr+1)}&=\big(\eta_{1,m}(itr)- \delta(itr)\big( Q_{m}\beta_{1,m}(|\hat h_{1,m}|^2\nonumber\\&+\xi_m|\hat h_{1,m}^{b}|^2|\hat h_{b,m}|^2-(2^{C_{min}}-1)(\sigma^2_\epsilon(Q_m\nonumber\\&\times(\beta_{1,m}+\beta_{2,m})+\xi_m)+I_{1,m}^{m'}+\sigma^2)\big)\big)^+\\
\eta_{2,m}^{(itr+1)}&=\big(\eta_{2,m}(itr)- \delta(itr)\big( Q_{m}\beta_{2,m}(|\hat h_{2,m}|^2+\xi_m\nonumber\\&|\hat h_{2,m}^{b}|^2|\hat h_{b,m}|^2-(2^{C_{min}}-1)(\varPi_{2,m}+\sigma^2_\epsilon(Q_m\nonumber\\&\times(\beta_{1,m}+\beta_{2,m})+\xi_m)+I_{2,m}^{m'}+\sigma^2)\big)\big)^+\\
\mu_{m}^{(itr+1)}&=\big(\mu_{m}(itr)-\delta(itr)\big(Q_{max}\nonumber\\ &-Q_m(\beta_{1,m}+\beta_{2,m})\big)\big)^+\\
\zeta_{m}^{(itr+1)}&=\big(\zeta_{m}(itr)-\delta(itr)\big(1-(\beta_{1,m}+\beta_{2,m})\big)\big)^+\\
\upsilon_{m}^{(itr+1)}&=\big(\upsilon_{m}(itr)-\delta(itr)\big(1-\xi_m\big)\big)^+
\end{align}
where (25)-(32) are iteratively updated until the selection criterion is satisfied.

\section{Numerical Results and Discussion}
Here we present the numerical results and their discussion. We compare the considered BC-NOMA V2XneT with the conventional NOMA V2XneT without BC. In this work, we calculate the system achievable energy efficiency (Mbpj) as the sum rate of NOMA V2XneT divided by the total power consumption. Unless stated otherwise the simulation parameters are set according to Table I.
\begin{figure}
\centering
    \includegraphics[width=0.46\textwidth]{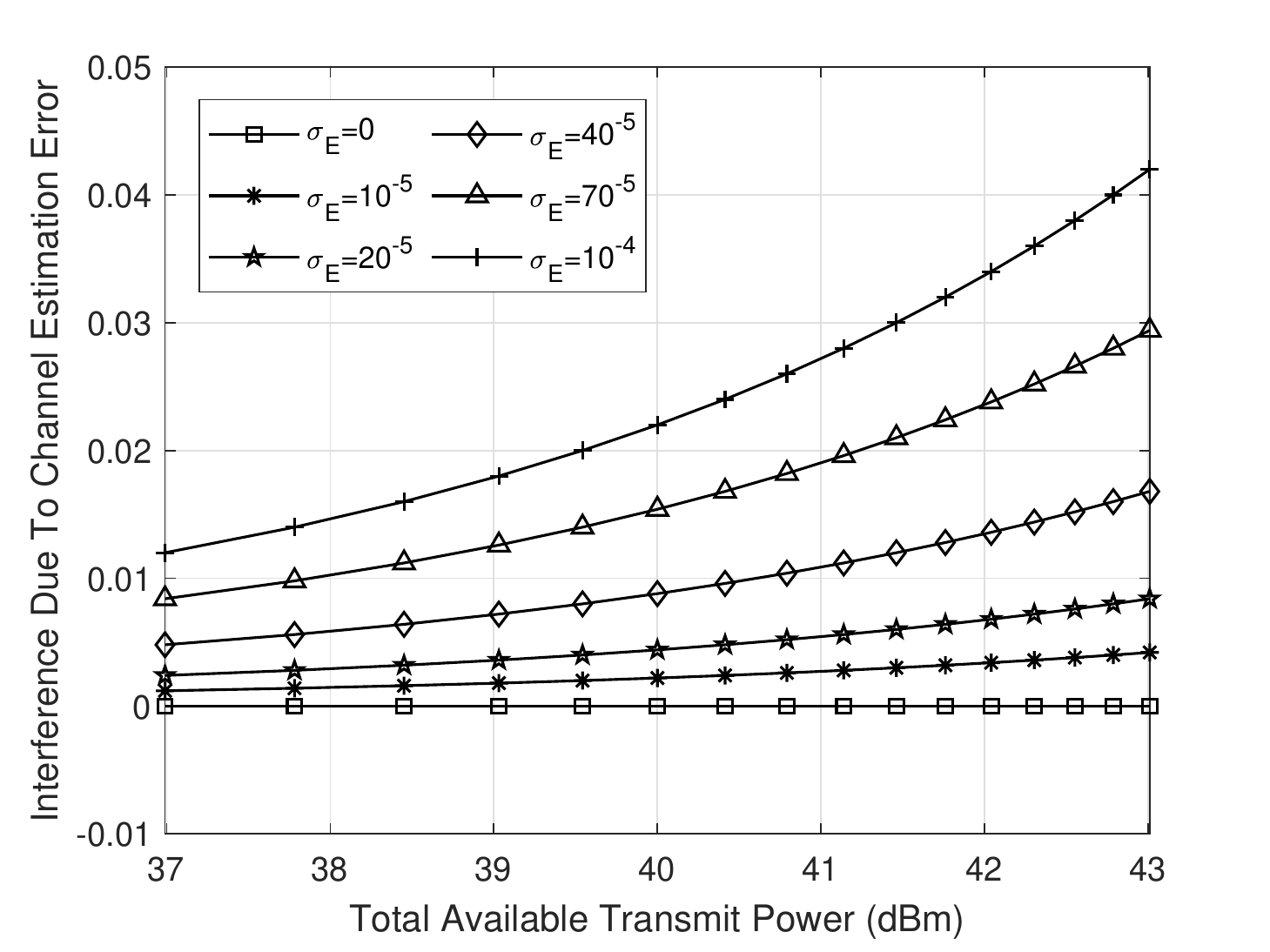}
    \caption{The effect of increasing transmit power on the interference due to channel estimation error.}
    \label{fig1d}
\end{figure}

\begin{table}[!t]
\centering
\caption{Simulation parameters}
\begin{tabular}{|c||c|} 
\hline 
Parameter & Value  \\
\hline\hline
Total power budget (BS+RSUs) & 45 dBm \\\hline
Maximum reflection coefficient $\xi$ & 1 \\\hline
Relay (RSUs) & Decode-and-forward \\\hline
Channel estimation error $(\sigma_\epsilon)$ & 0$\to$0.01\\\hline
Radius of BS & 50 meters\\\hline
Radius of RSU &  20 meters \\\hline
Channel realization & $10^3$ \\\hline
Minimum data rate $C_{\min}$ & 0.5 bps\\\hline
Pathloss exponent $(\zeta)$ & 4 \\\hline
Noise power density $\sigma^2$ & -170 dBm \\\hline
Bandwidth $B$ & 1 MHz \\\hline
Circuit power  & 5 dBm \\
\hline 
\end{tabular}
\end{table}

Fig. \ref{fig1d} shows the effect of imperfect CSI over the proposed BC-NOMA cooperative V2XneT by plotting interference due to the different values of channel estimation error versus the total available power of the system. We also plot perfect CSI where the value of $\sigma_E=0$, which means there is no interference and the devices can perceive accurate channel gain. We note that the interference is almost negligible for the small value of $\sigma_E$. However, by increasing the values of $\sigma_E$, the interference increases significantly as the available transmit power increases. It shows how important is the channel estimation in practical V2X scenarios. 

\begin{figure}
\centering
    \includegraphics[width=0.46\textwidth]{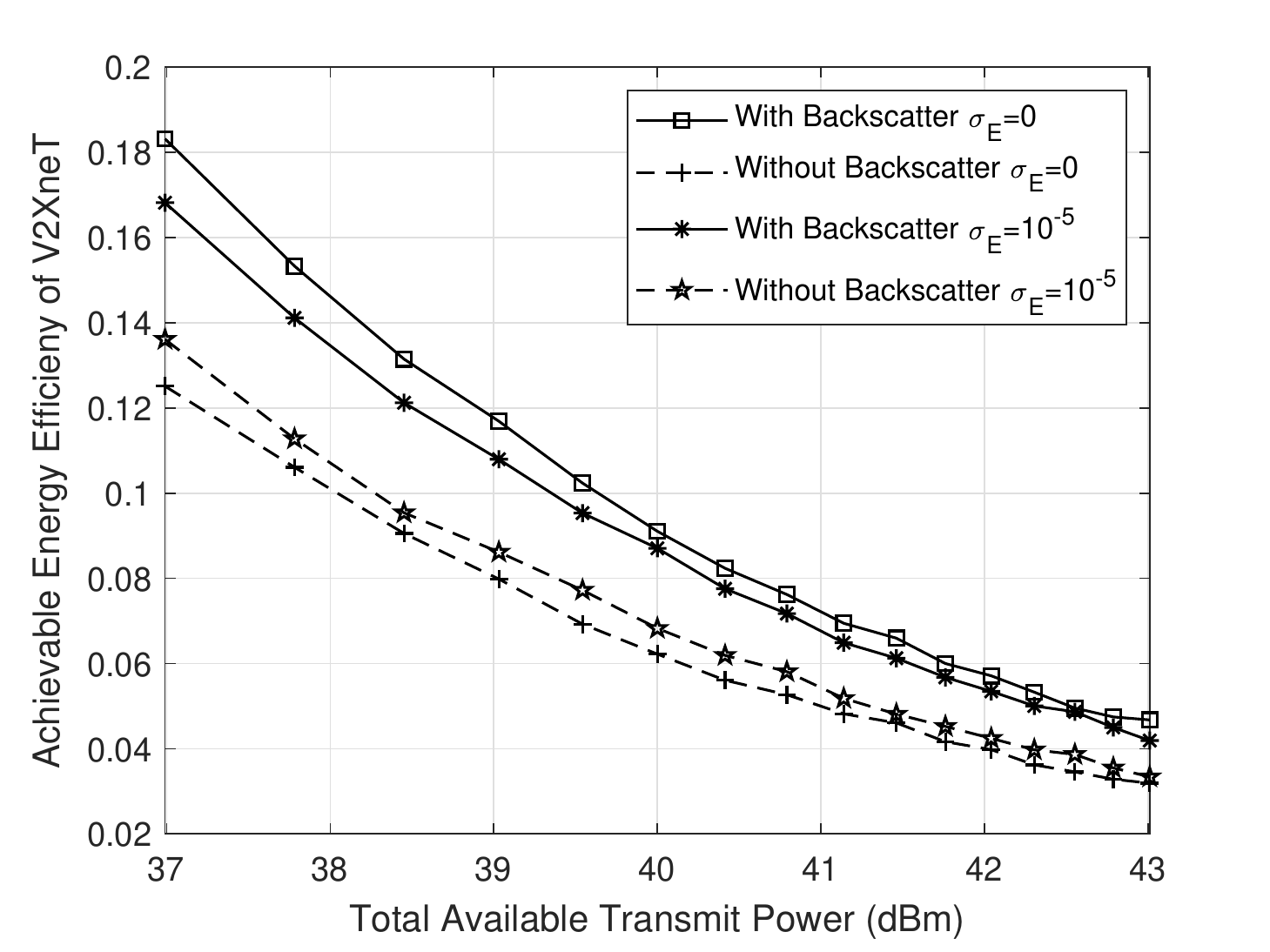}
    \caption{Total achievable energy efficiency of V2XneT versus available transmit power by varying $\sigma_{E}$.}
    \label{fig1a}
\end{figure}

Fig. \ref{fig1a} shows the achievable energy efficiency of V2XneT versus the available transmit power for different values of channel estimation error. Here we plot the achievable energy efficiency of both BC-NOMA and conventional NOMA V2XneT. We note an increase in the achievable energy efficiency of both frameworks against the increasing available power of the system. We can also observe that the curves follow the bell shape such that they increase initially when the available power of the system increase. Then it starts falling when it goes to the saturating point. However, the gap between the proposed BC-NOMA V2XneT and the conventional NOMA V2XneT is considerably large, showing the proposed framework's benefits.   

\begin{figure}
\centering
    \includegraphics[width=0.46\textwidth]{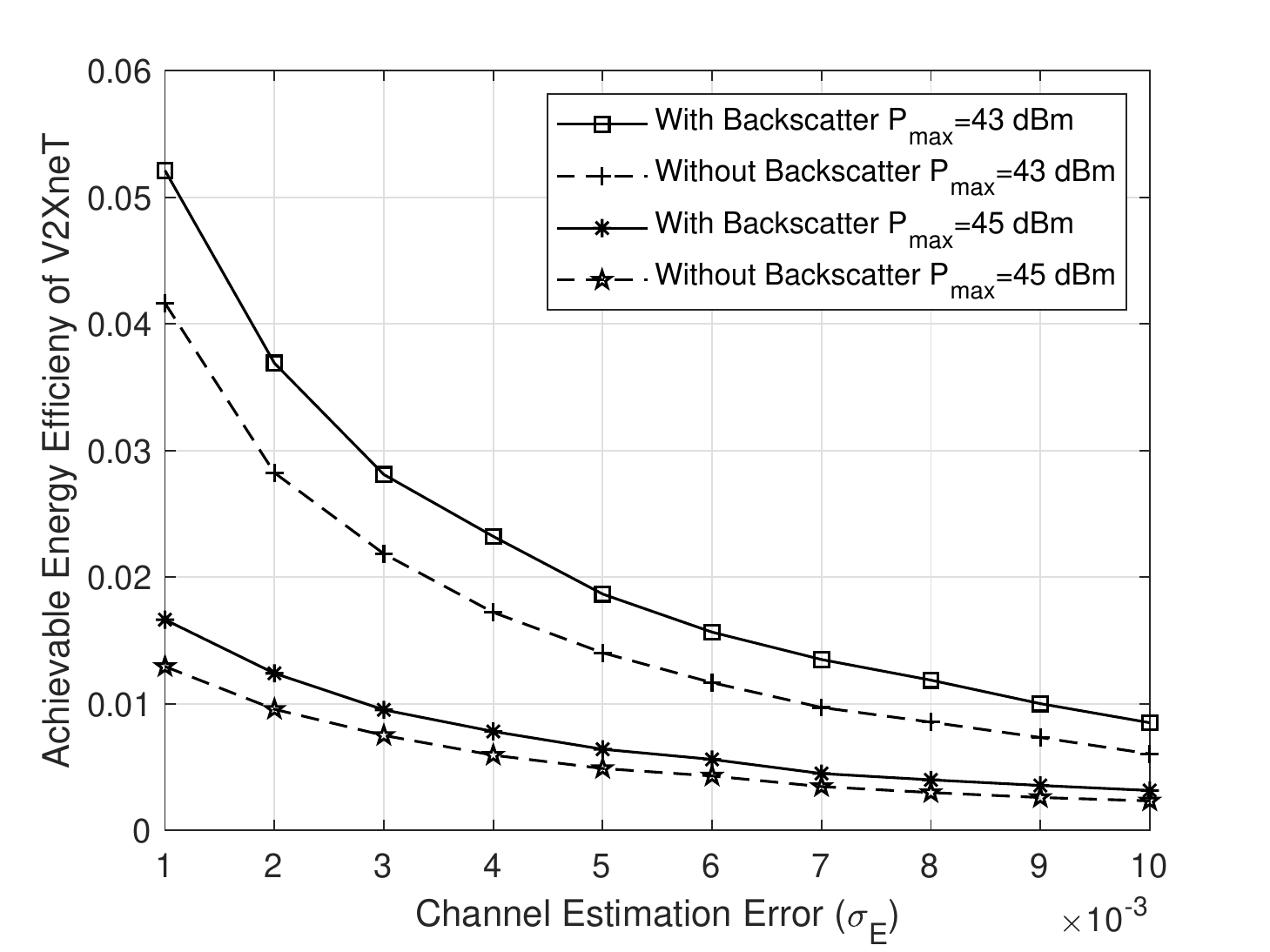}
    \caption{The effect of channel estmation error on the total achievable energy efficiency of V2XneT by varying $P_{max}$.}
    \label{fig1b}
\end{figure}
To further investigate the performance of the proposed BC-NOMA V2XneT, it is interesting to see the impact of varying channel estimation error on the achievable energy efficiency. Fig. \ref{fig1b} illustrates the system achievable energy efficiency against the varying values of $\sigma_E$ for different power budgets of V2XneT. As expected, the achievable energy efficiency of both frameworks decreases as the value of $\sigma_{E}$ increases. This is because when $\sigma_{E}$ increases, the interference is also increased, which decreases the sum rate of VEXneT. However, our BC-NOMA cooperative V2XneT performs better compared to the benchmark conventional NOMA V2XneT without BC. 

\begin{figure}
\centering
    \includegraphics[width=0.46\textwidth]{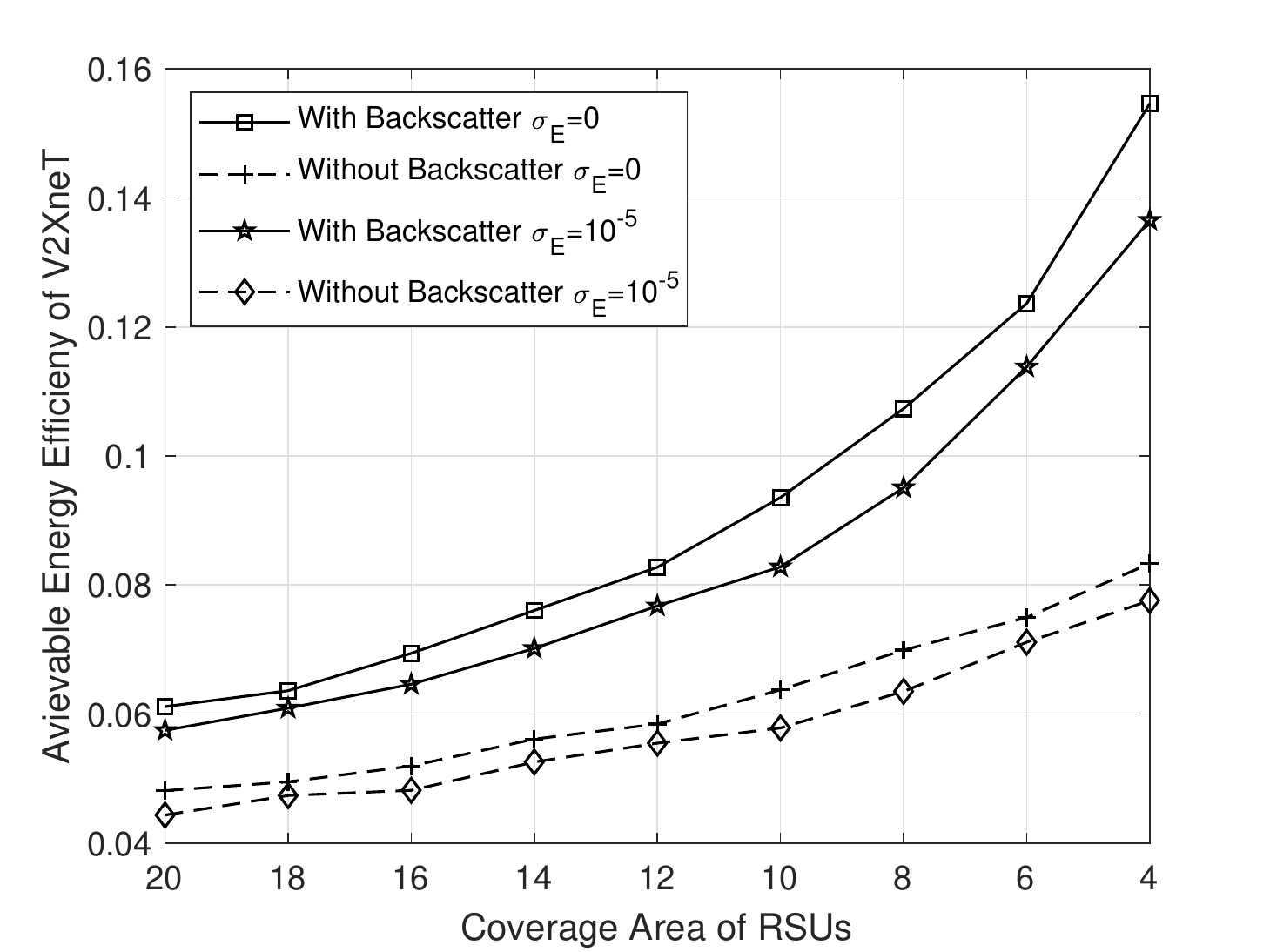}
    \caption{The effect of RSU coverage on V2XneT performance for different values of $\sigma_E$.}
    \label{fig1c}
\end{figure}
To show the performance of the proposed BC-NOMA V2XneT with the coverage area of RSUs, Fig. \ref{fig1c} investigates the achievable energy efficiency of BC-NOMA V2XneT against the varying coverage area of RSUs. We note that the achievable energy efficiency of both BC-NOMA and conventional NOMA V2XneT increases as the coverage area of RSU decreases. With small coverage, vehicles can achieve high data rate and require low transmit power for transmission. Another possible reason is the reflection of BD in the short range, resulting in comparatively high gain at vehicles. Another point worth mentioning here is that the proposed V2XneT outperforms the benchmark conventional V2XneT. 

\section{Conclusion}
For NextG V2XneT, NOMA and BC are the two promising technologies to improve energy and spectral efficiency. In this work, we have presented a new resource allocation framework for BC-NOMA V2XneT to improve the energy efficiency under the assumption of channel estimation error. In particular, the original problem has been decoupled into two sub-problems and solved by the iterative sub-gradient method. Simulation results have also been provided to demonstrate the benefits of the considered BC-NOMA V2XneT. It is important to note that the single BS and two RSUs have been the focus of this work. One exciting research direction is to extend this work to investigate the application of BC-NOMA in the dense heterogeneous scenario. 

\ifCLASSOPTIONcaptionsoff
  \newpage
\fi

\bibliographystyle{IEEEtran}
\bibliography{Wali_EE}

\end{document}